\documentclass[%
  reprint,
  superscriptaddress,
 amsmath,amssymb,
 aps,
 pre,
]{revtex4-1}

\usepackage{graphicx}
\usepackage{dcolumn}
\usepackage{bm}
\usepackage{physics}
\usepackage{hyperref}
\usepackage[mathlines]{lineno}
\usepackage{makecell} 

\begin{document}

\preprint{APS/123-QED}

\title{Void distributions reveal structural link between jammed packings  and protein cores}

\author{John D. Treado}
\affiliation{Department of Mechanical Engineering \& Materials Science, Yale University, New Haven, Connecticut 06520, USA}
\affiliation{Integrated Graduate Program in Physical \& Engineering Biology, Yale University, New Haven, Connecticut 06520, USA}

\author{Zhe Mei}%
\affiliation{Integrated Graduate Program in Physical \& Engineering Biology, Yale University, New Haven, Connecticut 06520, USA}
\affiliation{%
 Department of Chemistry, Yale University, New Haven, Connecticut 06520, USA
}%

\author{Lynne Regan}
\altaffiliation[current address: ]{Centre for Synthetic and Systems Biology, 
Institute for Quantitative Biology, Biochemistry, and Biotechnology, School of Biological Sciences, Edinburgh University, Scotland, UK}
\affiliation{
 Department of Molecular Biophysics \& Biochemistry, Yale University, New Haven, Connecticut 06520, USA
}%
\affiliation{%
 Department of Chemistry, Yale University, New Haven, Connecticut 06520, USA
}
\affiliation{Integrated Graduate Program in Physical \& Engineering Biology, Yale University, New Haven, Connecticut 06520, USA}

\author{Corey S. O'Hern}
\altaffiliation[email: ]{corey.ohern@yale.edu}
\affiliation{Department of Mechanical Engineering \& Materials Science, Yale University, New Haven, Connecticut 06520, USA}
\affiliation{Integrated Graduate Program in Physical \& Engineering Biology, Yale University, New Haven, Connecticut 06520, USA}
\affiliation{Department of Physics, Yale University, New Haven, Connecticut 06520, USA}
\affiliation{Department of Applied Physics, Yale University, New Haven, Connecticut 06520, USA}
\affiliation{Program in Computational Biology and Bioinformatics, Yale University, New Haven, Connecticut 06520, USA}

\date{\today}

\begin{abstract}
Dense packing of hydrophobic residues in the cores of globular
proteins determines their stability. Recently, we have shown that
protein cores possess packing fraction $\phi \approx 0.56$, which is
the same as dense, random packing of amino acid-shaped particles.  In
this article, we compare the structural properties of protein cores
and jammed packings of amino acid-shaped particles in much greater
depth by measuring their local and connected void regions. We find
that the distributions of surface Voronoi cell volumes and local
porosities obey similar statistics in both systems. We also measure
the probability that accessible, connected void regions percolate as a
function of the size of a spherical probe particle and show that both
systems possess the same critical probe size. By measuring the
critical exponent $\tau$ that characterizes the size distribution of
connected void clusters at the onset of percolation, we show that void
percolation in packings of amino acid-shaped particles and protein
cores belong to the same universality class, which is different from
that for void percolation in jammed sphere packings.  We propose that
the connected void regions of proteins are a defining feature of
proteins and can be used to differentiate experimentally observed
proteins from decoy structures that are generated using computational
protein design software. This work emphasizes that jammed packings of
amino acid-shaped particles can serve as structural and mechanical
analogs of protein cores, and could therefore be useful in modeling
the response of protein cores to cavity-expanding and -reducing
mutations.
\end{abstract}

\maketitle

\section{\label{sec:level1}Introduction}

\maketitle

A significant driving force in protein folding is the sequestration of
hydrophobic amino acids from solvent. Moreover, these buried amino
acids are densely packed in the protein
core~\citep{hydrophobic_rev}. In fact, the packing of core residues
has been linked directly to protein stability~\citep{wellpacked}. For
example, large-to-small amino acid mutations, which can increase
interior protein cavities, or voids, are known to destabilize proteins
when they are subjected to hydrostatic
pressure~\citep{cavities,cavities2,cavities3} and chemical
denaturants~\citep{Borgo1494,mutant1}. Understanding the connection
between dense core packing and voids is therefore crucial to
understanding the physical origins of protein stability and reliably
designing new protein structures that are
stable~\citep{rosetta-holes}. However, no such quantitative
understanding yet exists, and it is currently difficult to distinguish
computational protein designs that are not stable in experiments from
experimentally observed structures~\citep{capri}.

In previous studies~\citep{rcp_cores,unique_collrepack,sterics}, we
found, using collective side chain repacking, that the side chain
conformations of residues in protein cores (from a collection of
high-resolution protein crystal structures) are uniquely specified by
hard-sphere, steric interactions.  Moreover, we have shown that, when
considering hard-sphere optimized atomic radii, the core regions in
proteins possess the same packing fraction $\phi \approx 0.56$ as that
found in simulations of dense, random packings of purely repulsive,
amino acid-shaped particles. This result suggests that the packing
fraction of protein cores is determined by the bumpy and non-symmetric
geometries of amino acids, and not on the backbone or local secondary
structure.

However, materials that share the same packing fraction do not
necessarily possess the same internal structure.  In this article, we
characterize the void space in experimentally obtained and
computationally generated protein cores to further test the geometric
similarities between these two systems. We show below that dense
random packings of amino acid-shaped particles have the same local
packing fraction, void distribution, and percolation of connected void
space as protein cores, which indicates structural equivalence.

Our results suggest that the computationally generated packings can be
used as mechanical analogs of protein cores to predict their
collective mechanical response.  Further, our results emphasize the
connection between structurally arrested, yet thermally fluctuating,
protein cores and the jamming transition of highly nonspherical
particles~\citep{hypo}. Although the similarity between structural
glasses and proteins at low temperatures has been known for several
decades
\citep{glassy_proteins1,glassy_proteins2,glassy_proteins3,glassy_proteins4,glassy_proteins5},
prior computational studies have mainly focused on the transition from
harmonic to anharmonic conformational fluctuations on length scales
spanning the full protein.  In contrast, our studies identify key
structural similarities between jammed packings of amino acid shaped
particles and the cores of protein crystal structures.

This article is organized into four sections and three appendices. In
Sec.~\ref{sec:methods}, we describe the database of high-resolution
protein crystal structures that we use for our structural analyses and
the computational methods we use to generate jammed packings of amino
acid-shaped particles. We also outline two methods to measure the void
distribution in the two systems: a local measure of void space using
surface Voronoi tessellation, and a non-local or ``connected'' measure
of void space similar to that used by Kert\`esz~\citep{vperc1} and
Cuff and Martin~\citep{Cuff:2004aa}. In Sec.~\ref{sec:results}, we
compare the results of both the local and connected void measurements
for jammed packings of amino acid-shaped particles and protein cores
and find that both void measurements are the same for both systems. In
Sec.~\ref{sec:resultsA}, we show that the Voronoi cell volume
distributions in both systems are described by a $k$-gamma
distribution with similar shape factors $k$. In addition, we find that
the distribution of the local porosity ($\eta=1-\phi$) is the same for
protein cores and jammed packings of amino acid-shaped particles. In
Sec.~\ref{sec:resultsB}, we identify the percolation transition as a
function of the probe particle accessibility for the connected voids,
and find that protein cores and jammed packings of amino acid-shaped
particles share the same critical probe size that separates the
percolating and non-percolating regimes. We also investigate the
critical properties of this percolation transition, and show that it
is similar to void percolation of systems of randomly placed
spheres, but different from void percolation in jammed sphere
packings. In Sec.~\ref{sec:discussion}, we summarize our results,
discuss their importance, and identify future research directions. We
include three appendices with additional details of our computational
methods. In Appendix~\ref{appA}, we provide details for the
computational method we use to generate jammed packings of amino
acid-shaped particles. In Appendix~\ref{appB}, we discuss the
differences between protein cores in the Dunbrack 1.0 database, and
the core replicas we generate from jammed packings of amino
acid-shaped particles. In Appendix~\ref{appC}, we discuss the
differences between the connected void cluster size distributions in
jammed packings of spheres and amino acid-shaped particles and systems
containing randomly placed spheres.

\section{\label{sec:methods}Methods}

\begin{figure*}[t]
	\centering
    \includegraphics[width=0.975\textwidth]{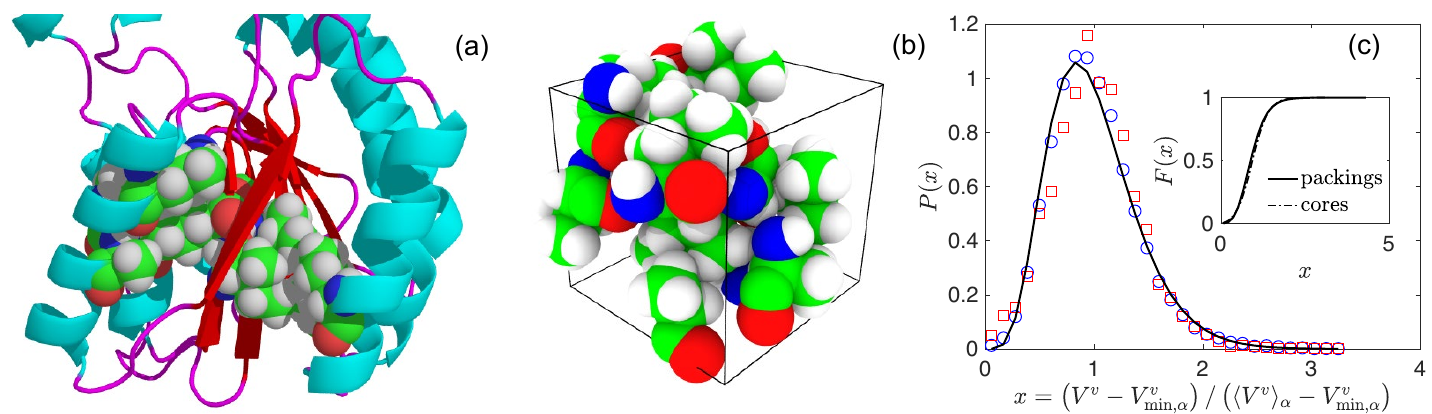}
\caption{(a) Core residues in an example globular protein (PDB code:
3F1L). Non-core regions are drawn using the ribbon representation,
and the $11$ core amino acids are drawn in all-atom
representation. (b) Jammed packing of the same $11$ core residues 
in (a). (c) The surface Voronoi cell volume $V^v$ distribution plotted 
as a function of $x = (V^v-V^v_{\text{min},\alpha})/(\langle V^v
\rangle_\alpha-V^v_{\text{min},\alpha})$ and 
fit to a $k$-gamma
distribution (black line) with $k = 6.06 \pm
0.08$ and $5.29 \pm 0.27$ for packings of amino acid-shaped particles 
(circles) and protein cores (squares), respectively. 
$\langle V^v
\rangle_\alpha$ is the average and $V^v_{\min,\alpha}$ is the minimum SV 
cell volume of residue type $\alpha$. The inset of (c) is
the cumulative distribution function $F(x)$ for the data in the 
main panel.}\label{svoro-dist}
\end{figure*}

To benchmark our studies of local and connected void regions, we use a
subset of the Dunbrack PISCES Protein Database (PDB) culling
server~\citep{dunbrack1,dunbrack2} of high-resolution protein crystal
structures. This dataset, which we will refer to as ``Dunbrack 1.0",
contains $221$ proteins with $< 50\%$ sequence identity, resolution
$\leq 1.0$~\AA, side chain $B$ factors per residue $\leq 30$ \AA$^2$
and $R$ factor $\leq 0.2$. We add hydrogen atoms to each protein
crystal structure using the \textsc{Reduce}
software~\citep{reduce}. To determine core amino acids, we calculate
the solvent accessible surface area (SASA) for each residue using the
\textsc{Naccess} software~\citep{naccess} with a $1.4$~\AA~water
molecule-sized probe~\citep{rSASA}. To compare the SASA for residues
with different sizes, we calculate the relative SASA (rSASA), which is
the ratio of the SASA of the residue in the protein context to that of
the residue \emph{outside} the protein context, along with the C$_\alpha$, C, and O atoms of the previous amino acid in the sequence and the N, H, and C$_\alpha$ atoms of the next amino acid in the sequence. We define core residues as those with rSASA $\leq
10^{-3}$.  We find similar results if the threshold for defining a
core residue is smaller, although there will be fewer ``core''
residues.  We showed in previous work that the local packing fraction
decreases significantly for residues with rSASA $> 0.05$~\citep{rSASA}.
(See Fig.~\ref{svoro-dist} (a) for an example core region in a protein 
from the Dunbrack 1.0 database.)

\begin{figure}
	\centering
\includegraphics[width=0.4\textwidth]{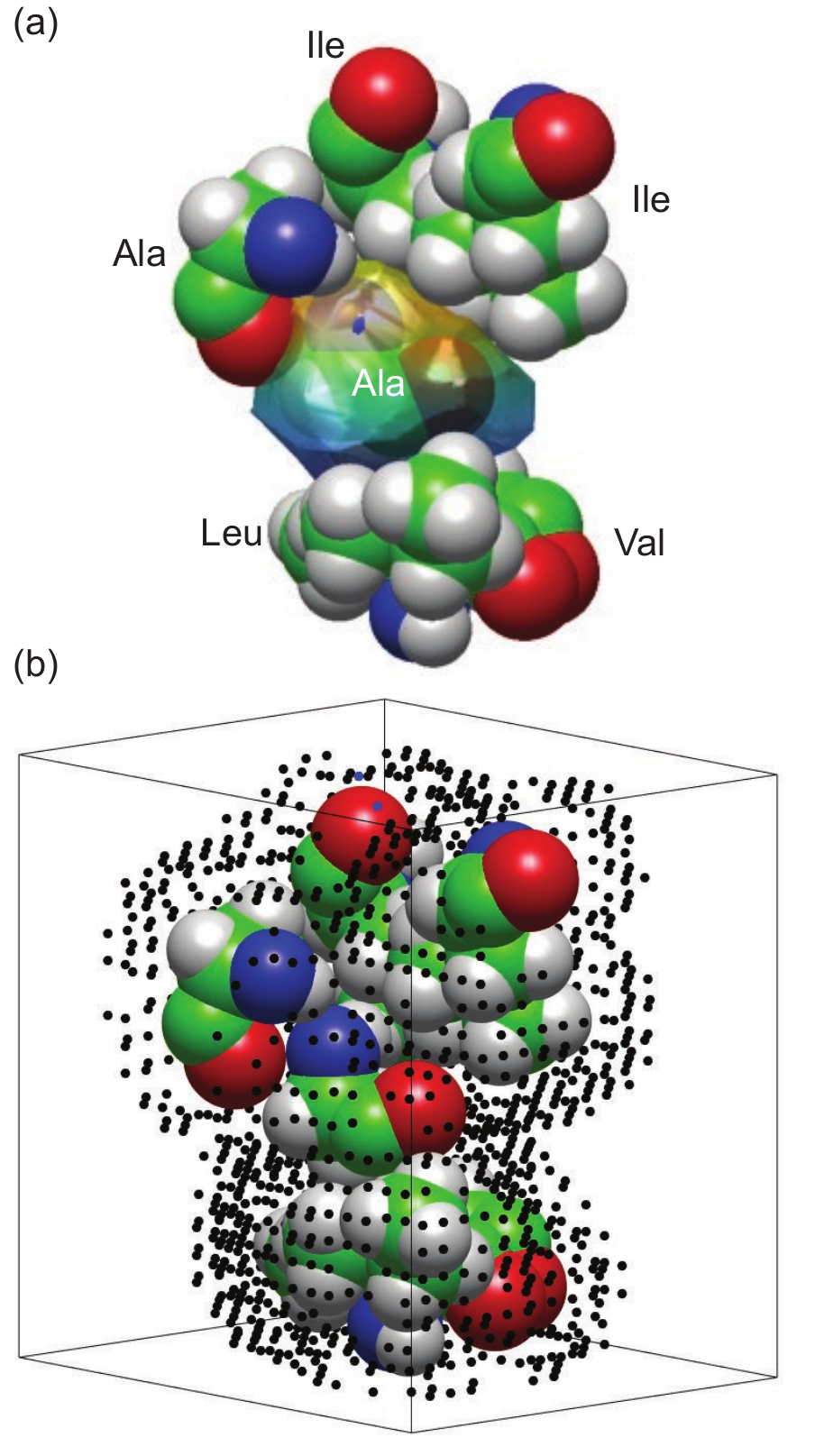}
\caption{Visualization of (a) local and (b) connected voids 
from the same computationally 
generated packing of $N=64$ amino acid-shaped particles. Only 
the central Alanine (Ala) with the neighboring Alanine, 2 Isoleucines (Ile), 
Leucine (Leu), and Valine (Val) are shown for clarity. The neighboring 
amino acids share at least one common surface Voronoi cell face with 
the central 
Ala. In (a), the central Ala is enclosed by its surface Voronoi cell. 
In (b), the connected void space is visualized using points on a grid. 
For clarity only $75\%$ of the points are shown,
and the grid spacing $(g = 0.7$\AA) is large compared to values
used in the text. In both (a) and (b), the atoms are colored as follows: 
C (green),
O (red), N (blue), and H (white). See
Fig.~\ref{cvoid_viz} for
visualizations of the connected void space throughout the entire
simulation domain.}
\label{void-viz}
\end{figure}

We will compare the structural properties of the cores of protein
crystal structures and jammed packings~\citep{epitome} of amino
acid-shaped particles.  In previous studies, we found that the packing
fraction of core regions in proteins is $\phi \approx 0.56$, which is
the same as that of jammed packings of purely repulsive amino
acid-shaped particles \textit{without} backbone
constraints~\citep{rcp_cores,pack-review}. Here, we
will focus exclusively on packing the hydrophobic residues: Ala, Leu,
Ile, Met, Phe, and Val.  The amino acid-shaped particles will include
the backbone atoms N, C$_\alpha$, C, and O, as well as all of the side
chain atoms, with the atomic radii given in
Ref.~\citep{rcp_cores}, which recapitulate the side chain dihedral
angles of residues in protein cores.  The packings of amino
acid-shaped particles contain mixtures of Ala, Leu, Ile, Met, Phe, and
Val residues, with each residue treated as a purely repulsive, rigid
body composed of a union of spherical atoms with fixed bond lengths,
bond angles, and side-chain and backbone dihedral angles taken from
instances in the Dunbrack 1.0 database.  

We choose which residues are
included in each packing using two methods. For method 1 (M1), we generate
${\cal C}=20$ jammed packings of the exact residues found in each
distinct protein core in the Dunbrack 1.0 database. For example, if
protein $X$ has a core with $R$ residues, we produce ${\cal C}$ jammed
packings of those exact $R$ residues. If $r$ of these $R$ residues are
not one of the hydrophobic residues we consider, these residues are
removed and a jammed packing is generated with the remaining $R-r$
residues. This method seeks to mimic the core size and amino acid
frequency distribution found in the Dunbrack 1.0 database. In method
2 (M2), we randomly select the hydrophobic residues with frequencies set by
that for each hydrophobic amino acid found in protein cores in the
Dunbrack 1.0 database. The frequencies are $0.29$ (Ala), $0.19$ (Leu),
$0.17$ (Ile), $0.05$ (Met), $0.07$ (Phe) and $0.23$ (Val). In method 2, 
the identities of the residues in the jammed packings only match those 
in protein cores on average. 

We now briefly describe the computational method for generating jammed
packings of amino acid-shaped particles. We use a pairwise, purely
repulsive linear spring potential to model inter-residue
interactions. Because the residues are rigid particles with each
composed of a union of spheres, we test for overlaps between residues
$\mu$ and $\nu$ by checking for overlaps between all atoms $i$ on
residue $\mu$ and all atoms $j$ on residue $\nu$, respectively. Note
that this potential is isotropic and depends only on the distances
between atoms on different residues. (See Eq.~\ref{linear} in
Appendix~\ref{appA}.)

We place $N$ residues with random initial positions and orientations
at packing fraction $\phi_0=0.40$ in a cubic simulation box with
periodic boundary conditions and then increase the packing fraction in
small steps $\Delta\phi$ to isotropically compress the system. After
each compression step, we relax the total potential energy using FIRE
energy minimization \citep{PhysRevLett.97.170201}. This method is
similar to a ``fast" thermal quench that finds the nearest local
potential energy minimum. We use quaternions to track the particle
orientations for each residue, as described in
Ref.~\citep{Rozmanov:2010aa}. If the total potential energy per
residue is zero after energy minimization, i.e. $U/N\epsilon <
10^{-8}$, where $\epsilon$ is the energy scale of the atomic
interactions, we continue to increase the packing fraction. If the
total potential energy per residue is nonzero, i.e. $U/N\epsilon \ge
10^{-8}$ and residues have small overlaps, we decrease the packing
fraction.  The packing fraction increment $\Delta\phi$ is halved each
time the algorithm switches from compression to decompression and vice
versa.  We terminate the packing-generation protocol when the residue
packings satisfy $10^{-8} < U/N\epsilon < 2\times10^{-8}$ and possess
a vanishing kinetic energy per residue (i.e. $K/N\epsilon <
10^{-20}$)~\citep{hypo}. (An example jammed packing of amino
acid-shaped particles is shown in Fig.~\ref{svoro-dist} (b) and
further computational details are included in Appendix~\ref{appA}.)

To measure the distribution of local voids in packings of amino
acid-shaped particles and protein cores, we use Voronoi tessellation,
which ascribes to each particle the region of space that is closer to
that particle than all other particles in the system. For residues,
which are highly non-spherical particles, we use a generalization of
the standard Voronoi tessellation known as the \emph{surface}- or
\emph{set}-Voronoi (SV) tessellation~\citep{M.-Schaller:2013aa}. This
tessellation partitions the empty space in the system using a bounding
surface for each residue. An efficient algorithm to
generate this tessellation is outlined in
Ref.~\citep{M.-Schaller:2013aa} and implemented using \textsc{Pomelo}
\citep{pomelo}. To construct the SV tessellation, consider a set of
$N$ particles with bounding surfaces $\qty{\partial K_\mu}$ for $\mu =
1,...,N$. The software approximates $\partial K_{\mu}$ by
triangulating points on the particle surfaces, and uses standard
Voronoi tessellation of the surface points to construct the SV cell
for each residue $\mu$. We find that using $400$ surface points per
atom, or $\approx 6400$ surface points per residue, gives an accurate
representation of the SV cell, which does not change significantly
as more surface points are added. An example SV cell from a packing of amino
acid-shaped particles is shown in Fig.~\ref{void-viz} (a). For an SV
cell with volume $V^v_\mu$ surrounding residue $\mu$ with volume
$v_\mu$, the local porosity is given by:
\begin{equation}\label{localporo}
	\eta_\mu = \frac{V^v_\mu-v_\mu}{V^v_\mu} = 1-\phi_\mu,
\end{equation}
where $\phi_\mu =v_\mu/V^v_\mu$ is the local packing
fraction. This quantity measures the local \emph{void} space associated with
each residue.

We also quantify the ``connected" void space shared between residues
in packings of amino acid-shaped particles and protein cores. To do
this, we implement a grid-based method similar to that described by
Kert\`esz~\citep{vperc1} and Cuff and Martin~\citep{Cuff:2004aa},
where the ``void space" is defined as the region of a system
accessible to a spherical probe particle with radius $a$. The geometry and
distribution of void space in a system is thus a function of $a$, the
residue positions ${\vec r}_{\mu}$, and bounding surfaces $\partial
K_{\mu}$. We define a cubic lattice with $G$ points in each direction
within the simulation domain, which gives a lattice spacing $g =
L/G$. For all lattice points $\vb{p}$, we define the set of void
points $\mathcal{V}$ to be all points that can accommodate a spherical probe
particle with radius $a$ without causing overlaps with any atoms. We
label all void points with a $1$, and all other points with a
$0$. After all grid points are labeled, we use the Newman-Ziff
algorithm~\citep{newman-ziff} to cluster adjacent similarly labeled
grid points. We consider all adjacent points on the nearest face,
edge, and vertex of a cube of points surrounding each lattice point
(i.e. next-to-next-to-nearest-neighbor counting with 26 possible
adjacencies for each point) when merging void clusters and implement 
periodic boundary conditions. A sketch of 
connected void lattice points in a subset of a packing of amino acid-shaped
particles is shown in Fig.~\ref{void-viz} (b).

When measuring void space in protein structures, we implement a
similar procedure, but we only consider voids in core residues. We
construct a box of dimension $L_x \times L_y \times L_z$ that
circumscribes each protein core, with the box just outside the
radii of core residues near the box edges. We pick a spherical probe particle of radius
$a$, and label the void space as all points that are (a) not contained
inside an atom, and (b) contained only within the union of the SV
cells of core residues. With these constraints, we only consider
connected void space specific to the core of the protein. We then use
the Newman-Ziff algorithm to merge void clusters, and repeat the
procedure for 100 different random protein orientations.

\section{Results}\label{sec:results}

\subsection{Local Void Analysis}\label{sec:resultsA}

We begin with an analysis of local voids associated with each amino 
acid in jammed packings of amino
acid-shaped particles and protein cores. We measure the distribution
of the SV cell volumes and show that the distributions in both systems
can be fit to a $k$-gamma distribution, which also describes Voronoi
cell distributions in jammed packings of
spheres~\citep{kgamma,Aste:2007aa},
ellipsoids~\citep{PhysRevX.6.041032},~attractive emulsion droplets
\cite{Jorjadze4286}, wet granular materials~\citep{Li:2014aa}, and
model cell monolayers~\citep{dp-paper}. The $k$-gamma distribution for 
the SV cell volume $V^v_\mu$ for each residue has
the form:
\begin{equation}\label{kg}
	P(x) = \frac{k^k}{\Gamma(k)}x^{k-1}\exp(-kx),
\end{equation}
where $x = \qty(V^v_\mu-V^v_{\min,\alpha})/\qty(\langle V^v_\mu \rangle_\alpha - V^v_{\min,\alpha})$, which sets the scale factor of the distribution 
to $1$. Here,
\begin{equation}
	\langle V^v_\mu \rangle_\alpha = \frac{1}{N_\alpha}\sum_{\mu=1}^{N_\alpha} V^v_\mu
\end{equation} 
is the average SV cell volume of residue type $\alpha$. The sum
involving $\mu$ is over all $N_{\alpha}$ residues of type $\alpha$ in
all packings, and $V^v_{\min,\alpha}$ is the minimum SV cell volume of
residue type $\alpha$.  We consider minima and averages for each 
residue type separately to account for the large differences 
in residue volumes; that is, each residue type $\alpha$, when considered individually, has a SV cell volume distribution described by Eq.~\eqref{kg}.

We measure the shape factor $k_\alpha$ for each residue type
$\alpha$ either by fitting the SV cell volume
distribution to Eq.~\eqref{kg} using Maximum
Likelihood Estimation (MLE), or by calculating 
\begin{equation}
	k_\alpha = \frac{\qty(\langle V^v_\mu \rangle_\alpha - V^v_{\min,\alpha})^2}{\left\langle \qty(V^v_\mu)^2 \right\rangle_\alpha - \left\langle V^v_\mu \right\rangle_\alpha^2 }.
\end{equation}
We obtain similar $k$-values using both methods. Although the values
of $k_\alpha$ depend on the type of amino acid $\alpha$, when we
average the values of $k_\alpha$ we recover the value of $k$ obtained
from fitting the combined distribution. We focus on the
distributions of SV cell volumes averaged over all hydrophobic
residues.

In Fig.~\ref{svoro-dist} (c), we show the SV cell volume distributions
$P(x)$ for packings of core amino acid-shaped particles modeled after specific protein cores (method M1) and for all core residues in the Dunbrack 1.0 database. We find that the distributions for these 
two systems are similar; 
both obey a $k$-gamma distribution [Eq.~\eqref{kg}] with similar shape 
parameters, $k = 6.06 \pm 0.08$ and $k = 5.29 \pm
0.27$, for core residues in the Dunbrack 1.0 database and packings 
of amino acid-shaped particles, respectively. As expected, the cumulative 
distributions $F(x)$ of the SV cell volumes for residues in protein cores and packings of amino acid-shaped particles are also nearly indistinguishable. 

\begin{figure}[t]
	\centering
    \includegraphics[width=0.475\textwidth]{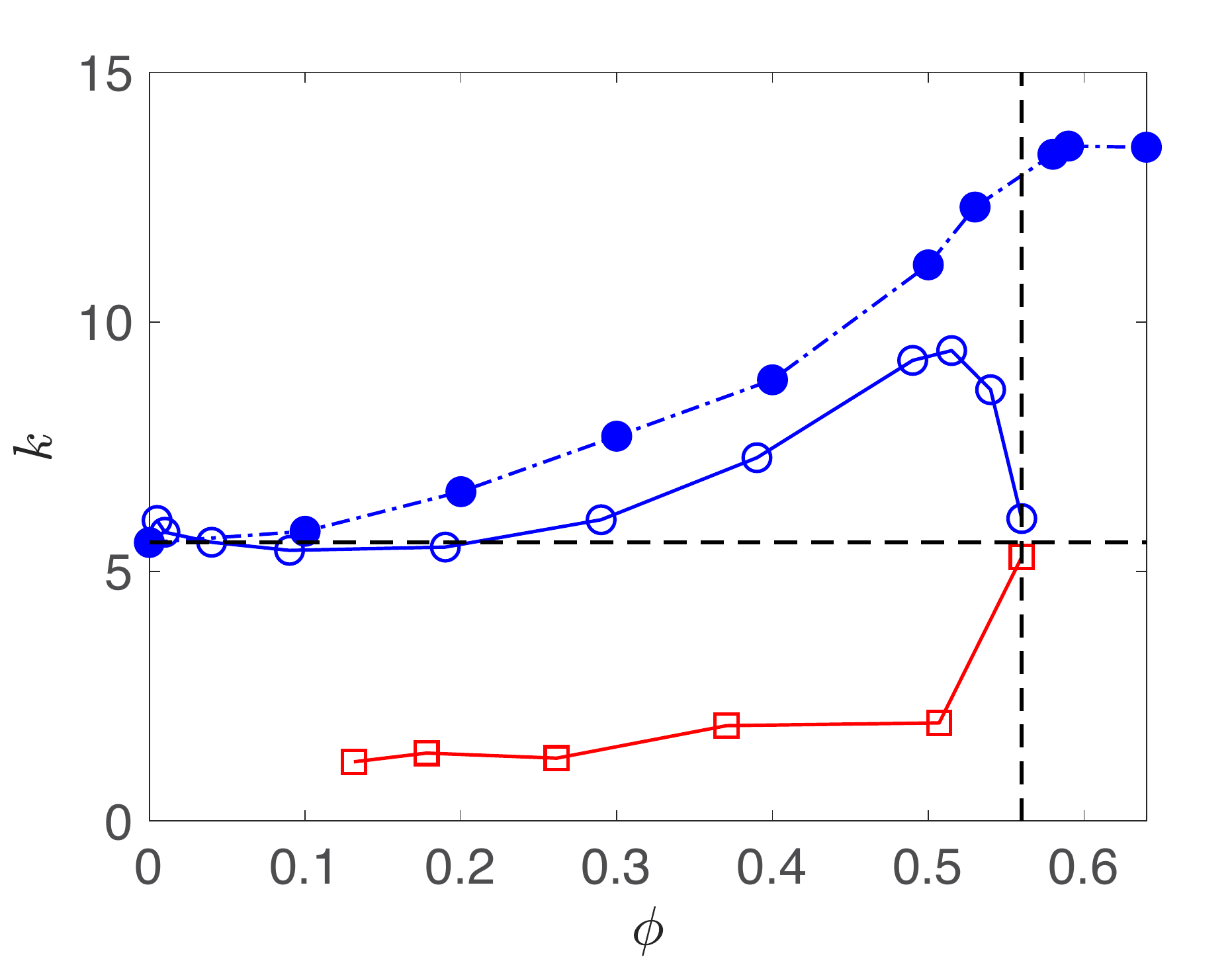}
\caption{The shape parameter $k$ for fits of the $k$-gamma distribution
[Eq.~\eqref{kg}] to the SV cell volume distributions $P(x)$ 
for packings of amino acid-shaped particles (open circles), 
monodisperse spheres (filled circles), 
and core residues in the Dunbrack 1.0
database (open squares) as a
function of packing fraction $\phi$. The dashed horizontal line
at $k = 5.59$ is the analytical value of the shape factor for
the Voronoi cell volume distribution of a random Poisson point 
process~\citep{poisson}, and the dashed vertical line at $\phi_J = 0.56$
is the  packing fraction for protein cores and jammed packings of amino 
acid-shaped particles.}
\label{svoro-phi}
\end{figure}
\begin{figure}[t]
	\centering
    \includegraphics[width=0.475\textwidth]{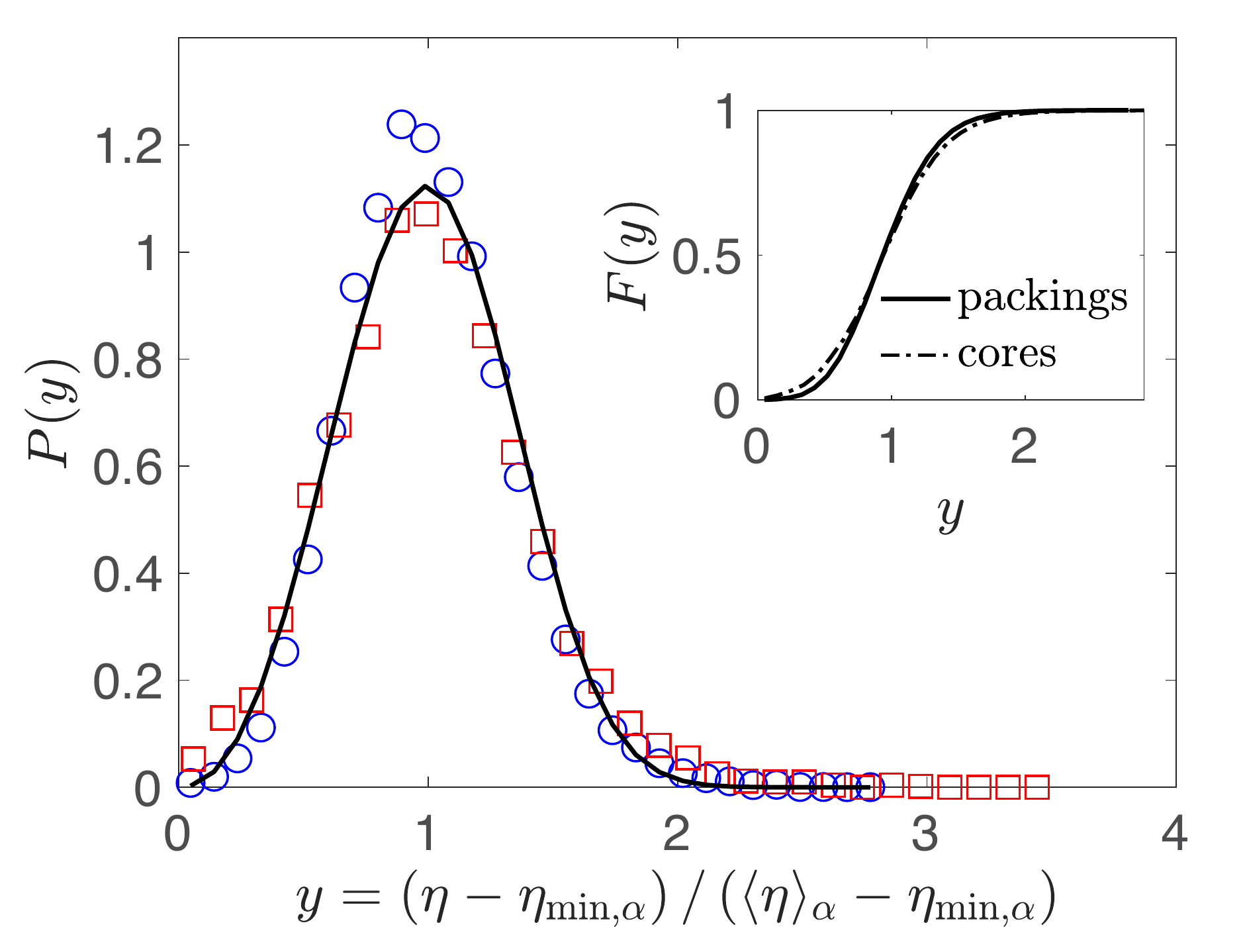}
\caption{Distribution of the scaled local porosity 
$y=(\eta-\eta_{\text{min},\alpha})/(\langle \eta
\rangle_\alpha-\eta_{\text{min},\alpha})$, where $\langle \eta
\rangle_\alpha$ is the average and $\eta_{\text{min},\alpha}$ is the 
minimum porosity of residue type $\alpha$, for packings of amino 
acid-shaped particles (circles) and residues in protein cores in the 
Dunbrack 1.0 database (squares). The solid line is a Weibull distribution 
with shape parameter $b \approx 3.2$ [Eq.~\eqref{wbl}]. The inset is the cumulative
distribution function $F(y)$ of the data in the main panel.}
\label{poro_plot}
\end{figure}

The strong similarity between the SV cell volume distributions
indicates that jammed packings of amino acid-shaped particles (at
$\phi_J \approx 0.56$) and protein cores possess the same
underlying structure. To better understand this result, in
Fig.~\ref{svoro-phi} we plot the shape parameter $k$ that describes
the form of the Voronoi cell volume distributions for packings of $N =
10^3$ monodisperse spheres (with $\phi_J \approx 0.64$) and of $N=64$
amino acid-shaped particles versus $\phi$. When $\phi \ll \phi_J$,
and the systems are sufficiently dilute, the Voronoi cell volume
distributions of the packings of monodisperse spheres and amino
acid-shaped particles resemble that for a random Poisson point
process~\citep{poisson} with $k \approx 5.6$. When $\phi \ll \phi_J$,
free volume is assigned randomly to each particle since
the particle positions are uncorrelated. However, as $\phi$ increases,
the $k$-values for packings of monodisperse spheres and amino acid-shaped
particles begin to grow, but at different rates, since the
particle geometry becomes important in determining the local free
volume. Near $\phi \simeq \phi_J$, the shape parameter plateaus at
$k\approx 13$ for packings of monodisperse spheres, but the shape
parameter decreases strongly to $k \approx 6$ for packings of amino
acid-shaped particles. This decrease in $k$ indicates a transition
from having the shape of the Voronoi cell volume distribution
determined by independent, weakly correlated particles (for $\phi
\lesssim \phi_J)$ to having the shape of the distribution determined
by bumpy, asymmetric amino acid-shaped particles (for $\phi \simeq
\phi_J$).

\begin{figure}[b]
	\centering
    \includegraphics[width=0.45\textwidth]{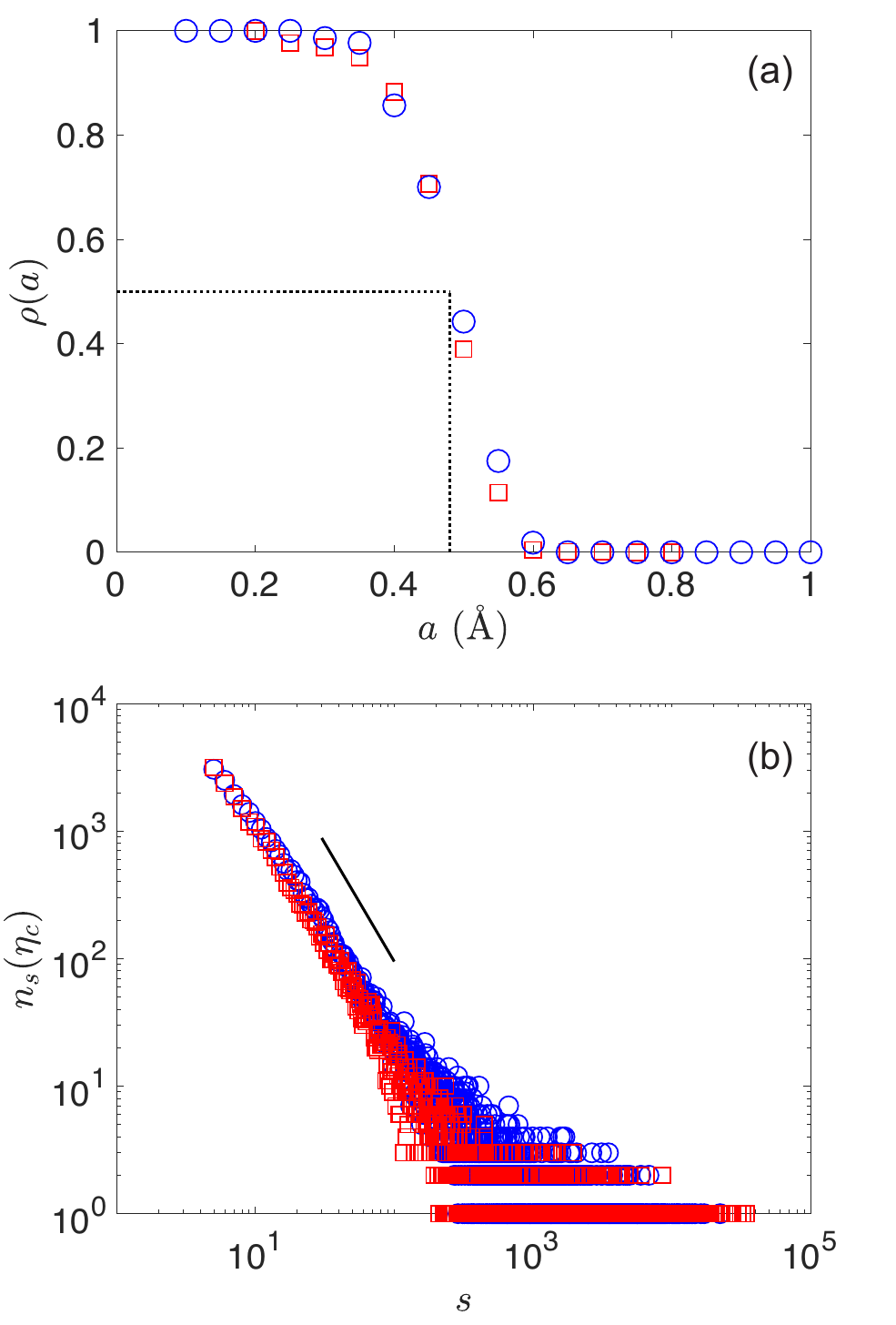}
\caption{(a) Percolation probability $\rho(a)$ plotted versus the
probe radius $a$ for protein cores from the Dunbrack 1.0
database (crosses) and clusters of core residues extracted from
static packings of $N = 64$ amino acid-shaped particles (circles). 
The horizontal and vertical dashed lines indicate the critical 
probe radius $a_c = 0.48$~\AA~that satisfies $\rho(a_c) = 0.5$. (b) Number
of connected void clusters $n_s$ with size $s$ at the percolating 
probe size $a_c$, which scales as $n_s(a_c) \sim s^{-\tau}$. 
The Fisher power-law exponent $\tau = 1.91 \pm 0.1$ and $1.8 \pm 0.09$ 
for protein cores from the Dunbrack 1.0 database (crosses) and 
representative clusters of core residues in 
packings of amino acid-shaped particles (circles),
respectively. The solid line has slope equal to 
-1.85. }
\label{crit_probe}
\end{figure}

We also calculate $k$ for the SV cell volume distributions for core
residues in the Dunbrack 1.0 database as a function of packing
fraction. For most of the range in $\phi$, $k \approx 2$, whereas $k
\gtrsim 5.6$ for packings of monodisperse spheres and amino acid-shaped
particles. In particular, $k$ does not equal the value for a random
Poisson point process ($k=5.6$) in the limit $\phi \ll \phi_J$ for
residues in protein cores. In protein cores, the backbone constraint
gives rise to correlations in the residue positions. However, as $\phi
\to \phi_J$, $k$ increases, reaching $k\approx 6$ when $\phi = \phi_J$.  This result shows that there is a fundamental change in the
SV cell distribution near the onset of jamming in protein cores. For
$\phi \lesssim \phi_J$, the backbone determines the shape of the
Voronoi cell volume distribution, whereas for $\phi \to \phi_J$, the
shapes of the amino acids determine the Voronoi cell volume
distribution.

We also compare the local porosity distributions for protein cores 
and packings of amino acid-shaped particles in Fig.~\ref{poro_plot}. 
We scale the porosity (as in Eq.~\eqref{kg}) by defining
\begin{equation}
	y = \frac{\eta_\mu - \eta_{\min,\alpha}}{\langle \eta_\mu \rangle_\alpha - \eta_{\min,\alpha}},
\end{equation}
where 
\begin{equation}
	\langle \eta_\mu \rangle_\alpha = \frac{1}{N_\alpha}\sum_{\mu=1}^{N_{\alpha}} \eta_\mu,
\end{equation}
and $\eta_{\min,\alpha}$ is the minimum porosity over all $N_{\alpha}$
core residues of type $\alpha$. Again, the porosity distributions $P(y)$
(and cumulative distributions $F(y)$) for residues in protein cores
and packings of amino acid-shaped particles are similar, but here $P(y)$
has the shape of a Weibull distribution with scale factor $\lambda=1$,
\begin{equation}\label{wbl}
	P(y) = by^{b-1}\exp(-y^b).
\end{equation}
where $b$ is the shape parameter of the Weibull distribution. 
   
The small differences in $P(x)$ and $P(y)$ between core residues in
protein crystal structures and packings of amino acid-shaped particles
can be explained by the small differences between the volumes of core
residues in crystal structures and in packings. The atoms on
neighboring amino acids interact differently for free amino acids in
packings versus backbone atoms in protein cores, which form covalent
and hydrogen bonds. Thus, we find that the volumes of residues in
protein cores have larger variances and smaller means than those in
packings of amino acid-shaped particles. Also, the overlaps between
covalently bonded backbone atoms that link adjacent residues slightly
decreases the mean SV cell volume, which gives rise to a larger
population of small SV cells and a small deviation between $P(x)$ for
residues in protein cores and in packings for small $x$ in
Fig. \ref{svoro-dist} (c).

\subsection{Connected Void Analysis}\label{sec:resultsB}

\begin{figure}[b]
	\centering
    \includegraphics[width=0.45\textwidth]{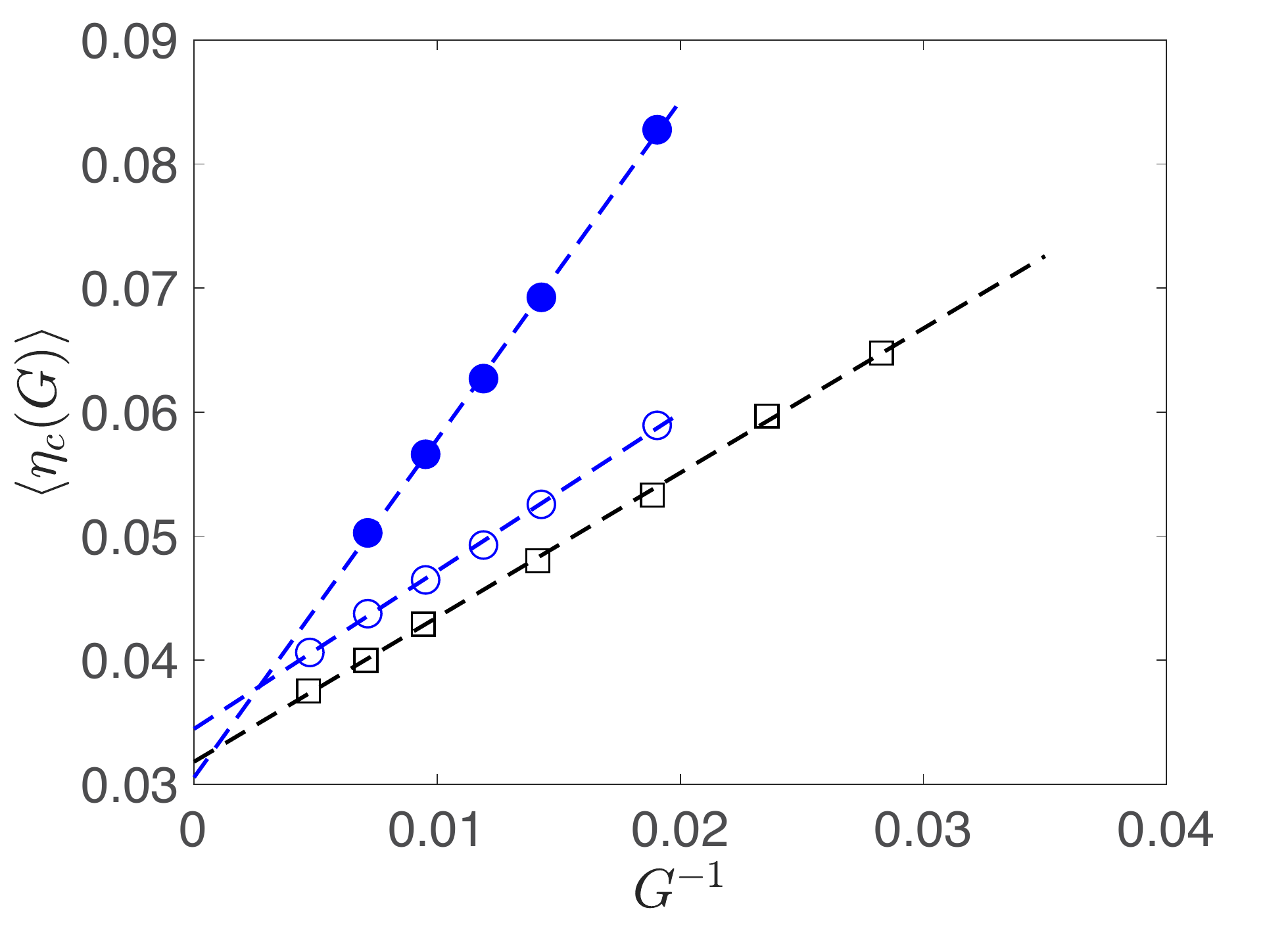}
\caption{Critical porosity $\langle \eta_c(G) \rangle$ using a lattice
with $G$ points along each dimension plotted versus $G^{-1}$ for jammed 
packings of $N = 10^3$ monodisperse spheres (filled circles) and of $N = 64$ 
amino acid-shaped particles (open circles), and $N=10^3$ randomly placed 
spheres (open squares). The dashed lines have vertical intercepts 
that indicate $\eta_c
\approx 0.0305$, $0.0318$, and $0.0343$ for the monodisperse sphere 
packings, packings of amino acid-shaped particles, and randomly 
placed spheres, respectively.}
\label{ctm_limit}
\end{figure}

\begin{figure*}
	\centering
    \includegraphics[width=0.9\textwidth]{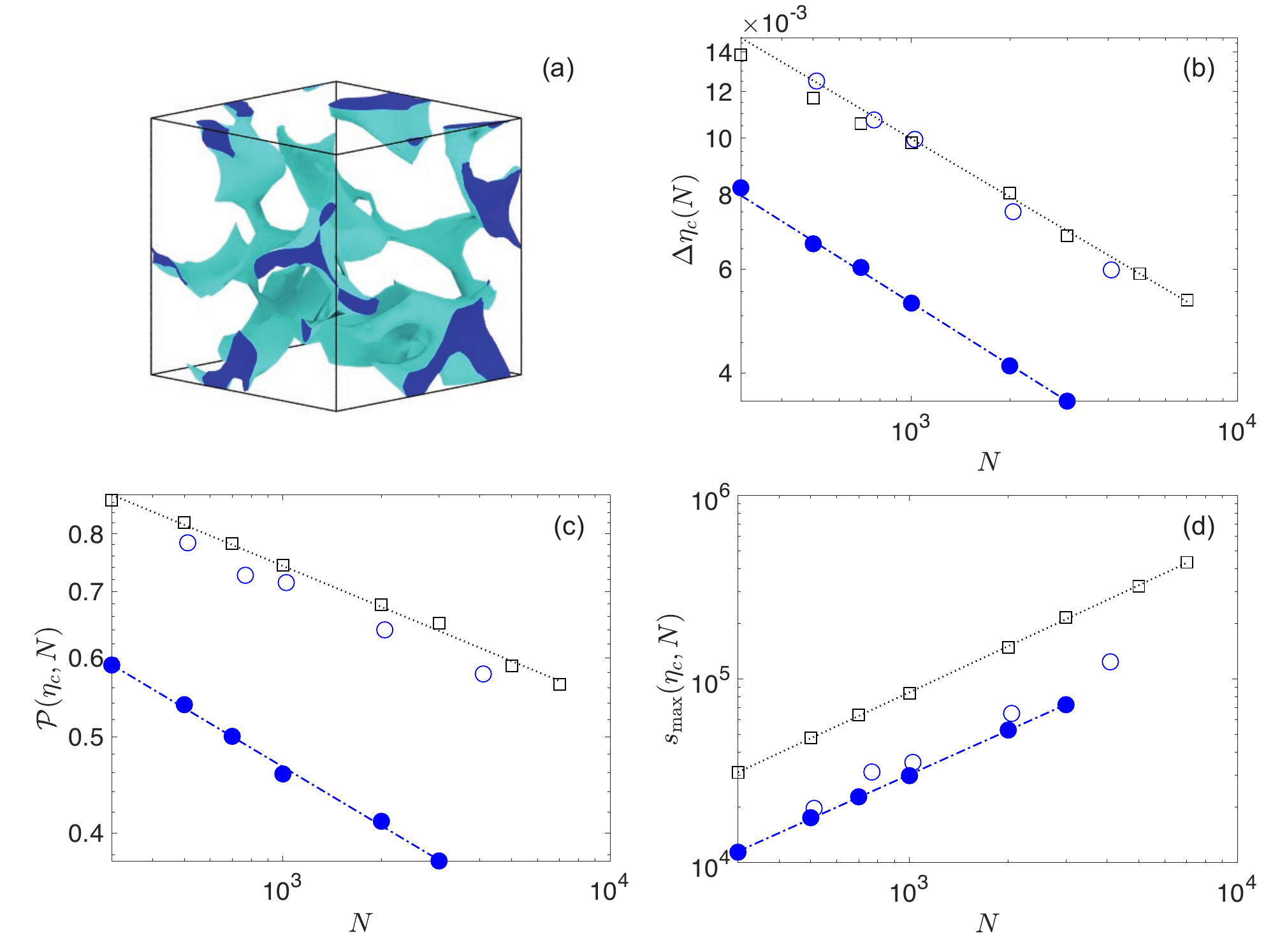}
\caption{(a) Visualization of the surface of a connected void region
(light domain) at the percolation threshold $\eta_c \approx 0.03$ in a
jammed packing of $N = 16$ amino acid-shaped particles (with $N_a =
256$ atoms). The dark domains are the ``inside'' of the void
region, which connects across the periodic boundaries. (b)-(d)
Finite-size scaling behavior for jammed packings of amino
acid-shaped particles (open circles), bidisperse spheres (filled
circles), and randomly placed spheres (open squares).  In (b), we
show that the standard deviation in the critical porosity scales as
$\Delta \eta_c (N) \sim N^{-1/d\nu}$, where $\nu$ is the correlation 
length exponent. The lines have slopes $-0.33$
(dotted line) and $-0.31$ (dot-dashed line). In (c), we show that the 
probability for a given site to be in the percolating void cluster 
at $\eta_c$ scales as ${\cal P}(\eta_c,N) \sim N^{-\beta/d\nu}$, where 
$\beta$ is the percolation strength scaling exponent.  The lines 
have slopes $-0.14$ (dotted line) and 
$-0.19$ (dot-dashed line).
In (d), we show that the maximum 
cluster size near the percolation onset scales as $s_{\rm max}(\eta_c,N) 
\sim N^{D/d}$, were $D$ is the fractal dimension. The lines have 
slopes $0.83$ (dotted line) and $0.82$ (dot-dashed line).  
}\label{finite_size_scaling}
\end{figure*}

We next quantify the distribution of ``connected'' void space that is
shared between residues. Using a grid-based method, we calculate the
volume of regions of connected void space as a function of the radius
$a$ of a spherical probe particle. As we increase $a$, the connected
void space transitions from highly connected throughout the system to
compact and localized with distinct void regions. We measure the
probability $\rho(a)$ of finding a percolating void region, where we
define percolation as the appearance of a cluster that spans one of
the system dimensions when the boundary is closed, and a cluster that
both spans, wraps around the boundary, and self-intersects when the
boundaries are periodic. We identify the critical probe radius $a_c$
by setting $\rho(a_c) = 0.5$. Because the definition of connected void
regions depends on the boundary condition, the value of $a_c$,
especially in systems as small as protein cores, is affected by the
boundary conditions. Thus, to calculate $\rho(a)$, we create packings
of amino acid-shaped particles with similar boundary conditions as
those in protein cores. From a packing of amino acid-shaped particles
with periodic boundary conditions ($N = 64$, method M2), we extract a
representative protein core of $R-r$ residues that all share at least
one SV cell face. We sample $R-r$ from the distribution of core sizes
$P(R)$ found in the Dunbrack 1.0 database. (See Fig.~\ref{fig:coresz}
in Appendix~\ref{appA}.)  The resulting packings have boundary
conditions similar to protein cores in the Dunbrack 1.0 database. We
then determine the connected void regions as a function of $a$ and
identify the critical probe size $a_c$ as shown in
Fig.~\ref{crit_probe} (a). We find the same critical probe size $a_c =
0.48 \pm 0.01$~\AA~for both protein cores and packings of amino
acid-shaped particles with similar boundary conditions. Note that this value of the critical probe radius is smaller than that of a water molecule, which is $\approx 1.4$ \AA, and thus the voids we consider here are not accessible by aqueous solvents. However, as we discuss below, this value of the probe radius corresponds to a critical point; we will exploit the behavior of the voids near this critical point to understand the geometric properties of the connected voids, and to differentiate between the voids in various systems. 

Thus, determining the connected void regions in protein cores is a
type of percolation problem. In lattice site percolation, sites on a
lattice in $d$ spatial dimensions are either occupied randomly with
probability $p$ or not occupied with probability $1-p$. At the
percolation threshold $p_c$, adjacent occupied sites form a
percolating cluster that spans the system and becomes infinite in the
large-system limit. Continuum percolation occurs in systems that are
not confined to a lattice. Both particle contact and void percolation
have been studied in randomly placed overlapping spheres~\citep{vperc1,
  vperc2, vperc3} and percolation of particle
contacts~\citep{pperc1,part-perc} has been studied in packings of
repulsive~\citep{tianqi-perc} and adhesive particles~\citep{corey-perc}.

In this article, we consider percolation of the void space accessible
to a spherical probe particle with radius $a$ in packings of spheres
and amino acid-shaped particles, as well as systems composed of
randomly placed spheres~\citep{vperc2,vperc3}. As the probe particle
radius is increased, the amount of space available to the probe is
restricted and the number of void lattice sites decreases. We define
an effective porosity $\eta$ as the ratio of the number of void
lattice sites to the total number of lattice sites $G^d$. We determine
the percolation threshold using a bisection method, where we begin
with two initial guesses for the percolation transition, $a_H$ and
$a_L$ with $a_H > a_L$, and iteratively check for percolation of void
sites at the probe radius $a = (a_H + a_L)/2$. We set $a_H = a$ if we
find a percolated cluster of void sites, and $a_L = a$ if we do not
find a percolated cluster. We terminate the algorithm when the
difference between successive values for $a_c$ are within a small
tolerance $\delta a = 10^{-8}$ \AA. Note that our use of a lattice of
points to measure the connected void region does not imply that our
model is a lattice model. The lattice is simply a tool to calculate
the connected void space volume~\citep{vperc1}.  Furthermore, in the
continuum limit (i.e. $G \rightarrow \infty$), we recover the critical
porosity $\eta_c \approx 0.03$ measured using Kerstein's
method~\citep{vperc2,vperc3} on systems of randomly placed
spheres~\citep{vperc4}. (See Fig.~\ref{ctm_limit}.) Since there is a
one-to-one mapping between $a$ and $\eta$, we will use $\eta$ as the
order parameter for continuum void percolation.

We now focus on the statistical properties of the connected void
regions in packings of spheres and amino acid-shaped particles. (See
Fig.~\ref{finite_size_scaling} (a) for an example connected void
region in packings of amino acid-shaped particles.) We first measure
the correlation length exponent $\nu$, where the correlation length
$\xi$ is defined as the average distance between two points in the
largest connected void cluster. Near $\eta_c$, $\xi$ diverges as
$\abs{\eta-\eta_c}^{-\nu}$.  Using finite-size
scaling~\citep{perc-theory}, we can write
\begin{equation}
	\eta_c(N)-\eta_c(\infty) \sim N^{-1/d\nu},
\end{equation}
where $\eta_c(\infty)$ is the percolation threshold in the
large-system limit and $N \sim L^d$. $\eta_c(N)$ is a random variable
with standard deviation $\Delta\eta_c(N)$, which will approach
$\eta_c(\infty)$ as $N\rightarrow\infty$. Thus, we make the ansatz
that
\begin{equation}\label{nu_eq}
	\Delta \eta_c(N) \sim N^{-1/d\nu},
\end{equation}
which can be used to measure $\nu$. (See
Fig.~\ref{finite_size_scaling} (b).) We also measure the probability
${\cal P}(\eta)$ that a given lattice site is part of the percolating void
cluster. Near $\eta_c$, the probability scales as ${\cal P}(\eta) \sim
\abs{\eta - \eta_c}^\beta$, where $\beta$ is the power-law exponent
that characterizes the ``percolation strength." The probability obeys 
finite size scaling,
\begin{equation}\label{beta_eq}
	{\cal P}(\eta,N) \sim N^{-\beta/d\nu}.
\end{equation}
Once we determine $\nu$ using Eq.~\eqref{nu_eq}, we can determine
$\beta$ from Eq.~\ref{beta_eq}. (See Fig.~\ref{finite_size_scaling}
(c).) We also expect $\beta$ and $\nu$ to satisfy the hyperscaling
relation,
\begin{equation}\label{hyperscaling_eq}
	D = d - \frac{\beta}{\nu},
\end{equation}
where $D$ is the fractal dimension of the percolating void cluster. The
fractal dimension is defined by
\begin{equation}\label{fracD_eq}
	s_{\text{max}}(\eta_c,N) \sim N^{D/d},
\end{equation}
where $s_{\text{max}}(\eta_c,N)$ is the number of sites contained in
the largest void cluster in the system at percolation onset. If $D =
d$, the largest void cluster is a compact, non-fractal
object. However, if $D < d$, the void cluster is
fractal~\citep{perc-text}. (See Fig.~\ref{finite_size_scaling} (d).)
We also measure the Fisher exponent $\tau$, defined by
\begin{equation}\label{fisher_eq}
	n_s(\eta_c) \sim s^{-\tau},
\end{equation}
where $n_s$ is the number of void clusters containing $s$ sites. We
measure this exponent for protein cores and random packings with
representative boundary conditions in Fig.~\ref{crit_probe} (b).

\begingroup
\squeezetable
\begin{table*}[t]
	\centering   
	\begin{tabular}{|c  c  c  c  c  c|}
    	\hline
        \hline
    	System & $\nu$ & $\tau$ & $D$ & $\beta$ & $d-\frac{\beta}{\nu}$\\
		\hline
    	residue packings (full) & $0.93 \pm 0.02$ & $1.15 \pm 0.06$ & $2.59 \pm 0.09$ & $0.40 \pm 0.0089$ & $2.57 \pm 0.013$\\
        residue packings (rep.) & $-$ & $1.8 \pm 0.08$ & $-$ & $-$ & $-$\\
        Protein cores, Dunbrack 1.0 & $-$ & $1.91 \pm 0.09$ & $-$ & $-$ & $-$\\        
        Mono. Spheres (jammed) & $1.07 \pm 0.08$ & $-$ & $2.46 \pm 0.04$ & $0.60 \pm 0.04$ & $2.44 \pm 0.06$\\
        Bidis. Spheres (jammed) & $0.96\pm 0.01$ & $-$ & $2.40 \pm 0.02$ & $0.56 \pm 02$ & $2.41 \pm 0.01$\\
        \hline
        Cubic Lattice & $0.91 \pm 0.03$ ($0.88$\footnote{Ref.~\cite{perc-theory}}) & $2.14 \pm 0.05$ ($2.18^{\text{a}}$) & $2.49 \pm 0.03$ ($2.53^{\text{a}}$) & $0.48 \pm 0.02$ ($0.42^{\text{a}}$) & $2.48 \pm 0.02$\\
        \makecell{Randomly Placed Spheres \\ (connected void method)} & $1.02 \pm 0.04$ & $1.1 \pm 0.05$ & $2.50 \pm 0.02$ & $0.42 \pm 0.01$ & $2.59 \pm 0.02$\\
        \makecell{Randomly Placed Spheres \\ (Voronoi vertex method)} & $1.00 \pm 0.05$ ($0.902\pm0.005$\footnote{Ref.~\citep{vperc3}}) & $-$ & $2.44 \pm 0.02$ & $0.5 \pm 0.01$($0.45\pm0.2$\footnote{Ref.~\citep{vperc2}}) & $2.50 \pm 0.03$\\
    	\hline
        \hline
	\end{tabular}
\caption{Table of critical exponents $\nu$, $\tau$, $D$, and $\beta$
for several models of void percolation. In the last column, we provide
the value for the hyperscaling relation,
$d - \frac{\beta}{\nu}$, which matches the fractal dimension $D$ if 
hyperscaling is satisfied. 
In the first four rows, we report the critical exponents
for packings of amino acid-shaped particles with periodic boundary
conditions (full) and boundary conditions representative of
protein cores (rep.). We also report the critical exponents for
void percolation in jammed packings of monondisperse (Mono.) and
bidisperse (Bidis.) spheres. In the last four
rows, we compare these results to those for void percolation in several 
systems that were 
studied previously. We report our measurements of the critical exponents for
site percolation on a cubic lattice, where only
nearest-neighbors are counted as adjacent sites. We also report the 
critical exponents for void percolation and Voronoi vertex
percolation in systems composed of randomly placed
spheres. Previously reported values of the exponents are given in
parentheses.}
\label{exp-table}
\end{table*}
\endgroup

In Table~\ref{exp-table}, we report our measurements for the critical
exponents $\nu$, $\tau$, $D$, and $\beta$ for void percolation (using
a spherical probe particle), as well as for $d = 3$ lattice site
percolation on a cubic lattice and void percolation in systems of
randomly placed spheres using two methods: the connected void
method described previously and the Voronoi vertex method
introduced by Kerstein~\citep{vperc2} and implemented by
Rintoul~\citep{vperc3}. Note that protein cores and representative
subsets of jammed packings of amino acid-shaped particles (denoted
``rep.")  are small systems with $N < 30$, and thus we cannot use
finite-size scaling to measure the critical exponents. We can,
however, measure the critical exponents for full packings of amino
acid-shaped particles (denoted ``full''), which mimic the geometric
properties of void clusters in protein cores.

We observe that across all models and methods studied, the correlation
length exponent $\nu \approx 0.9$-$1.0$ for void percolation. In
particular, $\nu \approx 0.93$ for packings of amino acid-shaped
particles is similar to that ($0.90$) for randomly placed
spheres~\citep{vperc3}. In addition, the fractal dimension $D \approx
2.4$-$2.6$ is similar for all models and methods for calculating void
percolation. We find that the percolation strength exponent $\beta <
0.5$ for randomly placed spheres and packings of amino acid-shaped
particles when using the connected void method, but $\beta >0.5$ for
packings of monodisperse and bidisperse spheres. (The bidisperse
systems include $N/2$ large and $N/2$ small spheres with diameter
ratio $d=1.4$.) This result suggests that the geometry of connected
void regions near percolation onset is most similar in packings of
amino acid-shaped particles and systems of randomly placed spheres.
In spite of the variations in the values of the exponents mentioned
above, the hyperscaling relation [Eq.~\eqref{hyperscaling_eq}] holds
for most systems.

The Fisher exponent $\tau$ [Eq.~\eqref{fisher_eq}] provides even
stronger evidence that the connected void regions in randomly placed
spheres and packings of amino acid-shaped particles are similar near
percolation onset.  For these two systems, $\tau \approx 1.1$ and
$1.15$. (See Table~\ref{exp-table}.) These values are distinct from
those for protein cores and residue packings with boundary conditions
similar to protein cores (i.e. $\tau \approx 1.8$ and $1.9$).  $\tau$
is sensitive to boundary conditions, and thus we expect these $\tau$ values to
differ. For lattice site percolation, $\tau \approx 2.14$ with
periodic boundary conditions, which is distinct from $\tau$ measured
in packings of amino acid-shaped particles and systems of randomly 
placed spheres with periodic boundary conditions.

\begin{figure*}[t]
	\centering
    \includegraphics[width=0.96\textwidth]{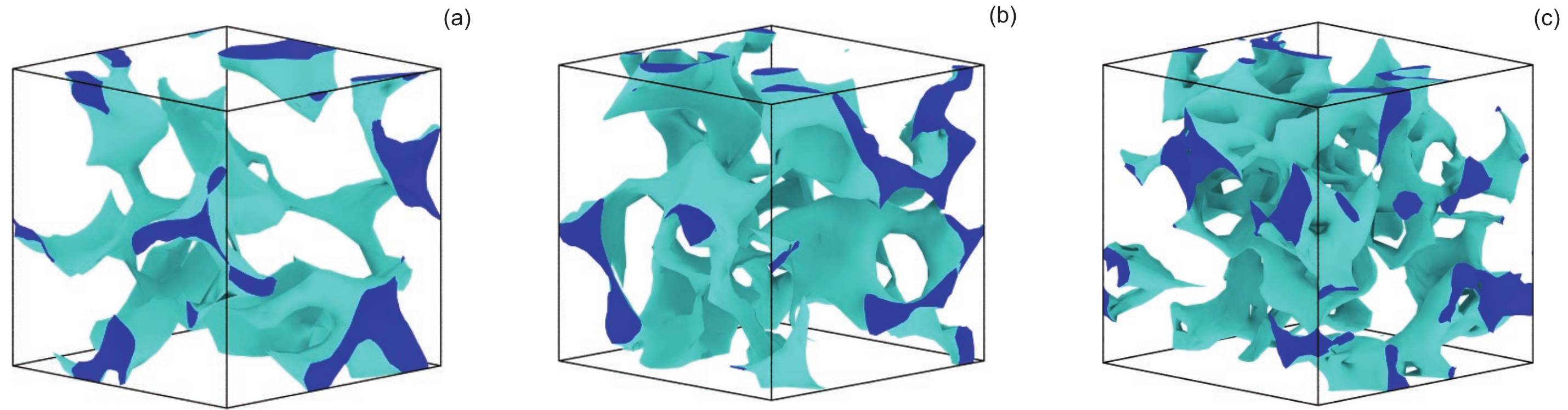}
\caption{Visualization of the surface of connected void regions (light domains)
near the percolation threshold in (a) a jammed packing of $N = 300$ amino-acid shaped particles, (b) 300 randomly placed
spheres, and (c) a jammed packing of $300$ bidisperse
spheres. The darker domains indicate the inside of the void region, 
which connect across the periodic boundaries.}\label{cvoid_viz}
\end{figure*}

We do not report values of $\tau$ for jammed packings of monodisperse
and bidisperse spheres, since we observe non-power-law behavior in the
cluster size distributions for these systems. As discussed in
Appendix~\ref{appC}, this behavior is most likely due to a residual
finite length scale at the percolation threshold. We also observe
non-power-law behavior in the cluster size distribution for void
percolation in randomly placed spheres using Kerstein's method, and do
not report a value for $\tau$ in Table~\ref{exp-table}. However, as
described in Appendix~\ref{appC}, the non-power-law behavior is most
likely due to system-size effects, which truncate the cluster size
distribution. Thus, we conclude that the critical exponent $\tau$ is
able to distinguish the geometries of connected void regions in
different systems. Moreover, our results suggest that the connected
void regions in packings of amino acid-shaped particles and systems of
randomly placed spheres belong to the same universality class, which
is distinct from that for jammed sphere packings. In
Fig.~\ref{cvoid_viz}, we show examples of the connected void surface
in packings of (a) amino acid-shaped particles, (b) randomly placed
spheres, and (c) bidisperse spheres. Qualitatively, the connected void
surfaces in systems of randomly placed spheres and amino acid-shaped
particles look similar, while the connected void surface in jammed
packings of bidisperse spheres looks different, with a
characteristic void size.

\section{Conclusions and Future Directions}\label{sec:discussion}

In this article, we analyzed local and connected void regions in
protein cores and in jammed packings of purely repulsive amino
acid-shaped particles and showed that these two systems share the same
void structure. We first investigated the surface-Voronoi (SV) cell
volume distributions and found that in both systems these
distributions are well-described by a $k$-gamma distribution with $k
\approx 6$. This $k$-value is much smaller than that ($k\approx 13$)
obtained for jammed sphere packings, which indicates that packings of
amino acid-shaped particles have a broader distribution of Voronoi
volumes. We also studied the SV cell volume distribution as a function
of the packing fraction, and found that only near the onset of jamming
do the SV cell distributions in protein cores and packings of amino
acid-shaped particles match.  In the dilute case $\phi \ll \phi_J$,
the local packing environment in protein cores is determined by the
backbone, whereas the local packing environment of packings of free
residues resembles a Poisson point process. At jamming onset, the
local packing environment is determined by the ``bumpy'', asymmetric
shape of amino acids, not the backbone constraints.

Using a grid-based method, we also measured the distribution of
non-local, connected voids in protein cores and jammed packings of
amino acid-shaped particles. We found that when we consider similar
boundary conditions in protein cores and jammed packings of amino
acid-shaped particles, the two systems also have the same critical
probe size $a_c$ (at which the accessible, connected void region spans
the system) and Fisher exponent $\tau$ (which characterizes the
scaling of the size of the void clusters near percolation onset). We
also compare the finite-size scaling results for void percolation in
packings of amino acid-shaped particles, in packings of monodisperse
and bidisperse spheres, and systems of randomly placed spheres.  We
find that void percolation in packings of amino acid-shaped particles
shares the same critical exponents as void percolation in randomly
placed spheres.  This result may also explain why the distribution of
SV cell volumes is similar for jammed packings of amino acid-shaped
particles and randomly placed spheres.

In future work, we will use jammed packings of amino acid-shaped
particles to understand the structural and mechanical response of
protein cores to amino acid mutations. We can assess the response in
two ways. First, we can prepare jammed packings of amino acid-shaped
particles that represent wildtype protein cores, substitute one or
more of the wildtype residues with other hydrophobic residues, relax
the ``mutant'' packing using potential energy minimization, and
measure the changes in void structure. We can also measure the
vibrational density of states (VDOS) in jammed packings that represent
the wildtype and mutant cores. The VDOS and the associated eigenmodes
can provide detailed information on how the low-energy collective
motions change in response to mutations. There are several advantages
for calculating the VDOS in jammed packings of amino acid-shaped
particles.  For example, in jammed packings, only hard-sphere-like
steric interactions are included.  In contrast, molecular dynamics
force fields for proteins typically include many terms in addition to
those that enforce protein stereochemistry, which makes it difficult
to determine the interactions that control the collective
motions. Studying jammed packings of amino acid-shaped particles also
decouples the motions of core versus surface residues.

Studies of the VDOS in jammed packings of amino acid-shaped particles
will also shed light on the protein ``glass'' transition, where the
root-mean-square (rms) deviations in the atomic positions switch from
harmonic to anharmonic behavior~\citep{glassy_proteins4} in globular
proteins near $T_g \approx 200$ K~\citep{glassy_proteins5}. We will
investigate the vibrational response of jammed packings of amino
acid-shaped particles to thermal fluctuations.  In particular, we will
measure the Fourier transform of the position fluctuations and
determine the onset of anharmonic response.
 
In addition, our analysis of void distributions in protein cores will
provide new methods for identifying protein decoys, which are
computationally generated protein structures that are not observed
experimentally. However, it is currently difficult to distinguish between real structures and decoys. For example, in the most recent Critical Assessment of Protein Structure Prediction (CASP12), researchers were given a set of target sequences, and were tasked with predicting the structures of those sequences using a variety of methods~\citep{casp1}. Each group was allowed to submit 5 structures per target sequence; when tasked with assessing which of their submissions were the most accurate, only 3 groups out of 31 had $> 50\%$ success at identifying the most accurate structure~\citep{casp2}. The average success rate was $30\%$, just slightly better than guessing at random. Thus, assessing the viability of computationally-designed structures is an incredibly difficult task.

Since the structure of void regions in the cores of protein crystal structures
is the same as that found in packings of amino acid-shaped particles,
the properties of void regions can serve as a benchmark for ranking
computationally designed protein structures. Recent studies have
suggested that protein decoys~\citep{rosetta-holes} possess local
packing fraction inhomogeneities that are not present in protein
crystal structures. We propose that detailed characterizations of the void 
space, using the methods described here, will be a sensitive metric than 
can be used to assess a variety of protein designs. 

\section*{Acknowledgements}
The authors acknowledge support from NIH training Grant, Grant No.
T32EB019941 (J.D.T), the Raymond and Beverly Sackler Institute for
Biological, Physical, and Engineering Sciences (Z.M.), and NSF Grant
No. PHY-1522467 (C.S.O.). This work also benefited from the facilities
and staff of the Yale University Faculty of Arts and Sciences High
Performance Computing Center. We thank J. C. Gaines for providing the
code to analyze cores in protein crystal structures and Z. Levine for 
helpful comments on this research.

\appendix

\section{Packing-generation Protocol}
\label{appA}

As described in Sec.~\ref{sec:methods}, we generate jammed packings of
amino acid-shaped particles using successive small steps of isotropic
compression or decompression with each step followed by potential
energy minimization. Each residue was modeled as a rigid union of
spheres with fixed bond lengths, bond angles, and dihedral angles.
The purely respulsive forces between residues were obtained by
considering small overlaps between atoms on different residues,
and then applying these forces to the center-of-mass of each residue,
which gives rise to translational and rotational motion. Forces
between atoms $i$ and $j$ on distinct residues $\mu$ and $\nu$ were
calculated using ${\vec F}_{ij}^{\mu\nu} = -{\vec \nabla}
U\qty(r^{\mu\nu}_{ij})$, with the pairwise, purely repulsive 
linear spring potential energy,
\begin{equation}
\label{linear}
	U\qty(r^{\mu\nu}_{ij}) = \frac{\epsilon}{2}\qty(1 - \frac{r^{\mu\nu}_{ij}}{\sigma^{\mu\nu}_{ij}})^2\Theta\qty(1 - \frac{r^{\mu\nu}_{ij}}{\sigma^{\mu\nu}_{ij}}).
\end{equation}
In Eq.~\ref{linear}, $\epsilon$ is the characteristic energy scale of
the repulsive interactions, $\sigma^{\mu\nu}_{ij} =
(\sigma^{\mu}_i+\sigma^{\nu}_j)/2$, $r^{\mu\nu}_{ij} =
|{\vec r}^{\mu}_j-{\vec r}^{\nu}_i|$ is the separation between atoms 
$i$ and $j$ on distinct residues $\mu$ and $\nu$, and $\Theta$ is the  
Heaviside step function that sets the potential energy to zero when 
atoms $i$ and $j$ are not in contact. Note that this pair potential 
reduces to a hard-sphere-like interaction in the limit of small atomic 
overlaps~\citep{pack-review}. The total potential energy $U$ is given by
\begin{equation}
U = \sum_{\nu<\mu}\sum_{i,j} U\qty(r^{\mu\nu}_{ij}).
\end{equation}
We use the velocity-Verlet algorithm to integrate the translational
equations of motion for each particle's center of mass, and a
quaternion-based variant of the velocity-Verlet method described
in Ref.~\citep{Rozmanov:2010aa} to integrate the rotational equations
of motions for each residue. 

\begin{figure}
	\centering
    \includegraphics[width=0.475\textwidth]{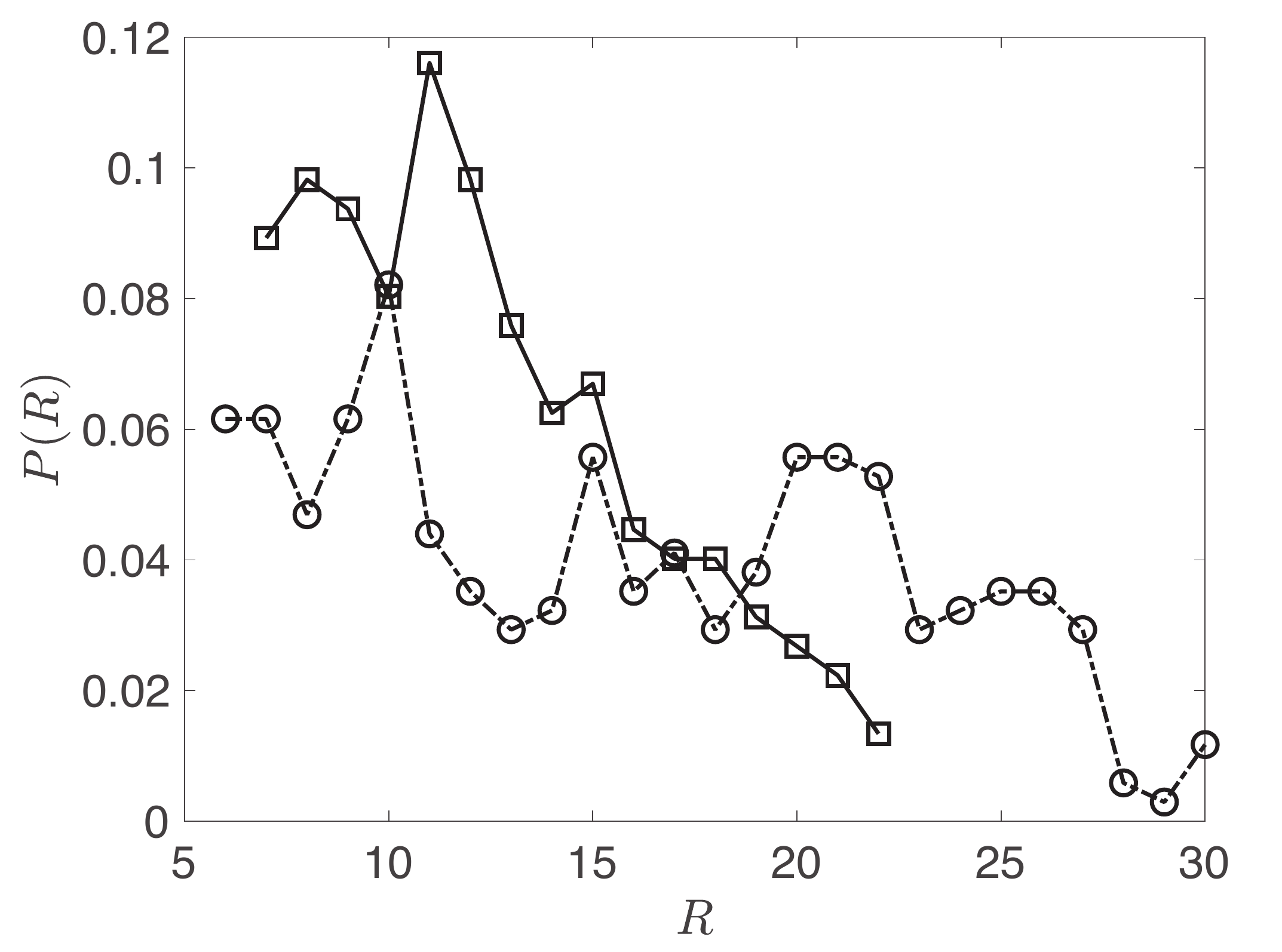}
\caption{Distribution of the number of core residues $P(R)$ in the
Dunbrack 1.0 database, before (circles) and
after (squares) pruning non-hydrophobic residues
from the core replicas as described in Sec.~\ref{sec:resultsB}. The mean number 
of residues 
before pruning is $\langle R \rangle \approx 16$, and after pruning 
is $\langle R-r
\rangle \approx 12$.}\label{fig:coresz}
\end{figure}
\begin{figure}[t]
	\centering
    \includegraphics[width=0.475\textwidth]{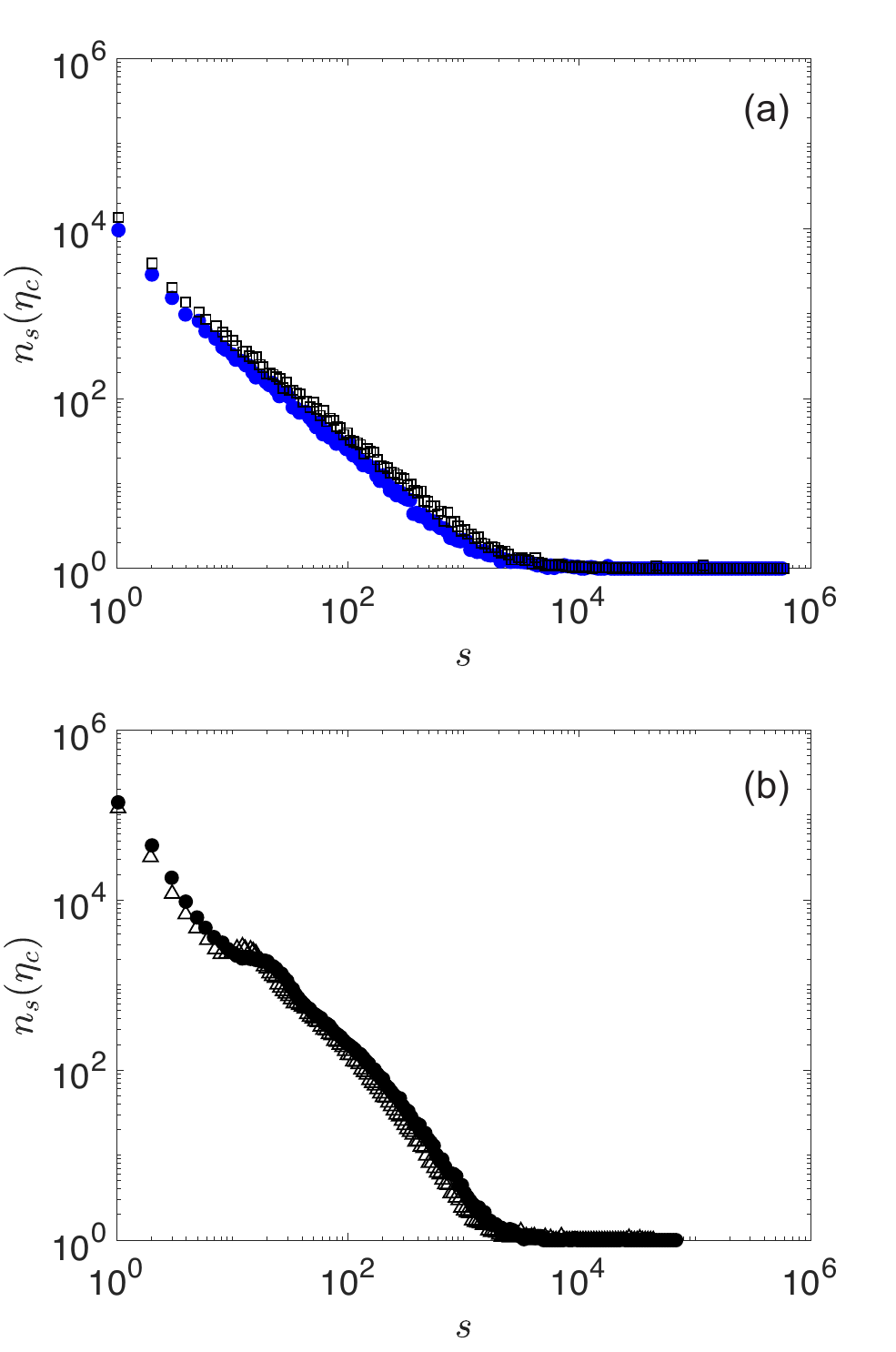}
\caption{Void cluster size distribution at percolation onset $n_s(\eta_c)$ 
for (top) packings of amino acid-shaped particles (filled 
circles) and randomly placed spheres
(open squares) and (bottom) jammed packings of monodisperse (filled circles) 
and bidisperse spheres (open triangles). In both panels, we use the 
connected void method to measure $n_s(\eta_c)$. The distributions in the 
top panel have well-defined power-law decay for $s < 10^4$, $n_s(\eta_c) \sim s^{-\tau}$ with 
exponent $\tau \approx 1.1$, whereas the distributions in the bottom panel are not strict 
power-laws over the same range of $s$. The plateau regions at large $s$ are due
to finite-size effects.}\label{fisher1}
\end{figure}
\begin{figure}
	\centering
    \includegraphics[width=0.45\textwidth]{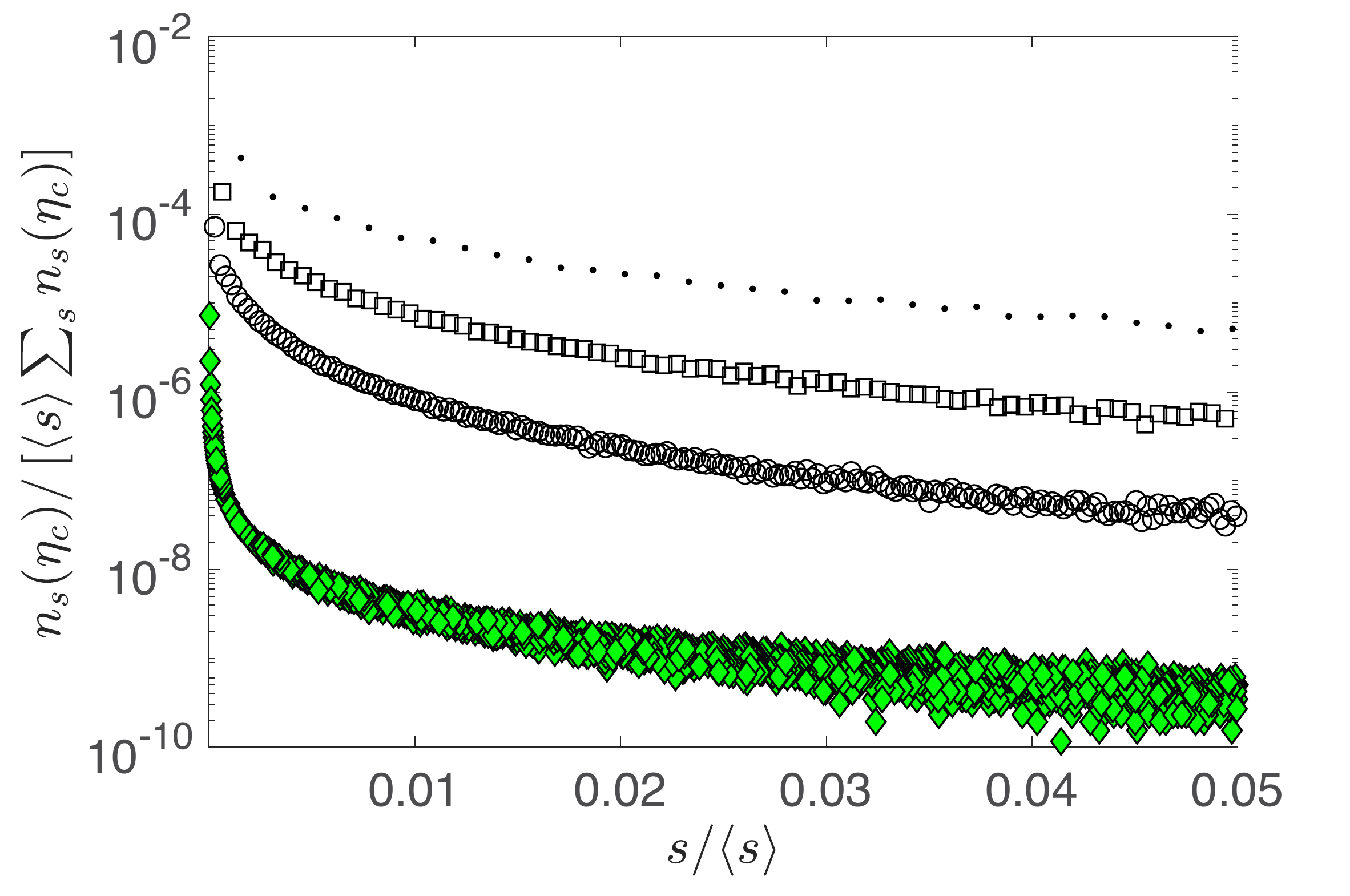}
\caption{Cluster size distribution at percolation $n_s(\eta_c)$ versus 
$s/\langle s\rangle$, normalized such
that each curve has unit area, for void percolation though
randomly placed spheres using Kerstein's method with $N = 300$ (dots), 
$10^3$ (squares), and $10^4$ (open circles). We also show the normalized 
$n_s(\eta_c)$ calculated using the connected void method for 
$N = 7\times 10^3$ randomly placed sphere (filled diamonds).}
\label{vertex_tau}
\end{figure}

To simulate isotropic compression, we scale all lengths in the system
(except the box edges) at each iteration $m$ by the scale factor 
\begin{equation}
	\alpha = \qty(\frac{\phi_m + \Delta \phi_m}{\phi_m})^{1/3},
\end{equation}
where $\phi_m$ is the packing fraction and $\Delta \phi_m$ is the
packing fraction increment at iteration $m$. This process uniformly
grows or shrinks all atoms, and thus the packing fraction satisfies
$\phi_{m+1} = \phi_{m}+\Delta\phi_m$. After each compression or
decompression step, we use the FIRE algorithm
\citep{PhysRevLett.97.170201} to minimize the potential energy in the
packing. The packing fraction increment is halved each time the total
poential energy switches from zero (i.e.  $U/N\epsilon < 10^{-8}$) to
nonzero or vice versa. We terminate the packing-generation algorithm
when the total potential energy per residue satisfies $10^{-8} <
U/N\epsilon < 2 \times 10^{-8}$ and the kinetic energy per
residue is below a small threshold, $K/N \epsilon < 10^{-20}$. We set
the initial values of the packing fraction and packing fraction
increment to be $\phi_0 = 0.4$ and $\Delta \phi_0 = 10^{-3}$, but
our results do not depend sensitively on these values.

\section{Protein Core Size Distribution}
\label{appB}

In this Appendix, we show the distributions of the number of core
residues in protein crystal structures from the Dunbrack 1.0
database. (See Fig.~\ref{fig:coresz}.) As described in
Sec.~\ref{sec:methods}, we define protein cores as clusters of
residues that all share at least one SV cell face with other residues 
in the cores, and every atom in
each residue has an rSASA $\leq 10^{-3}$. In Method M1 for generating
jammed packings of amino acid-shaped particles, we create ${\cal C}$
replicas of each protein core with the specific $R-r$ residues found in that
core, where $R$ is the number of core residues and $R-r$ is
the number of Ala, Ile, Leu, Met, Phe, and Val core residues.
Before pruning non-hydrophobic residues, the average core size
is $\langle R \rangle \approx 16$ residues, and $\langle R-r
\rangle \approx 12$ after pruning.

\section{Measurement of the Fisher Exponent $\tau$}\label{appC}

In this Appendix, we explain the differences we observe in the Fisher
exponent $\tau$ for different systems. In systems of randomly placed
spheres and in jammed packings of amino acid-shaped particles, the
distribution of void cluster sizes at percolation onset $n_s(\eta_c)$
has a well defined power-law decay, as shown in Fig.~\ref{fisher1}
(a).  Non-power-law decay in the void cluster size distribution, as
displayed in Fig.~\ref{fisher1} (b) for jammed sphere packings, may be due to
the existence of multiple important length scales in the system. The
typical form of Eq.~\eqref{fisher_eq} at any porosity $\eta$ is~\citep{perc-theory}
\begin{equation}
\label{new}
	n_s = s^{-\tau}\exp(-s/s_\xi),
\end{equation}
where $s_\xi$ is the number of sites in a cluster with correlation
length $\xi$. In systems where $\xi$ is the only length scale, $s_\xi
\rightarrow \infty$ as $\eta \rightarrow \eta_c$ and Eq.~\ref{new}
reduces to Eq.~\eqref{fisher_eq}. However, if there is another
intrinsic length scale in the system that is still relevant at the
void percolation transition, it is not necessarily true that $s_\xi
\rightarrow \infty$. $s_{\xi}$ can remain finite, and add an
exponential tail to $n_s(\eta_c)$. Indeed, this behavior is what we
find for the connected void size distribution in jammed sphere
packings. The ``kink'' in $n_s(\eta_c)$ in Fig.~\ref{fisher1} (b)
indicates that $s_\xi \approx 10$. .

This second length scale is most likely set by the neareast neighbor
distances between particles. Qualitatively, if the nearest-neighbor
distance between particles is a $\delta$-function (or a set of
$\delta$-functions, in the case of polydisperse spheres), there are a
limited number of local cavities in the system. In particular, there
can be small, particle-scale voids that persist even even at the
percolation threshold. However, in packings of amino acid-shaped
particles and in systems of randomly placed spheres, there are a wide
range of inter-particle distances, and a continuous range of local
cavity sizes that can form. In Fig.~\ref{cvoid_viz}, we show that the
void regions are well-connected for jammed packings of amino
acid-shaped particles and randomly placed spheres, while the void
regions have a characteristic cavity size for jammed sphere packings
at percolation onset. Thus, there is a well-defined
Fisher exponent $\tau$ in jammed packings of amino acid-shaped
particles and randomly placed spheres, but not in jammed monodisperse 
and bidisperse sphere
packings.

\newpage
\bibliography{voids}

\end{document}